\input{aipcheck}

\documentclass[
    ,final            % use final for the camera ready runs
  ]
  {aipproc}

\layoutstyle{6x9}

\begin{document}

\title{Lag-luminosity relation in gamma-ray burst X-ray flares}


\author{R. Margutti}{
  address={raffaella.margutti@brera.inaf.it}
}

\begin{abstract}
In strict analogy to prompt pulses, X-ray flares observed by \emph{Swift}-XRT in long Gamma-Ray Bursts define a lag-luminosity 
relation: $L_{\rm{p,iso}}^{0.3-10 \,\rm{keV}}\propto t_{\rm{lag}}^{-0.95\pm0.23}$. The lag-luminosity 
is proven to be a fundamental law extending $\sim$5 
decades in time and $\sim$5 in energy. This is direct evidence that GRB X-ray flares and prompt 
gamma-ray pulses are produced by the same mechanism. 
\end{abstract}

\maketitle

%%%%%%%%%%%%%%%%%%%%%%%%%%%%%%%%%%%%%%%%%%%%
%% MAINMATTER
%%%%%%%%%%%%%%%%%%%%%%%%%%%%%%%%%%%%%%%%%%%%

\section{Introduction}
The presence of X-ray flares superimposed to Gamma-Ray Bursts (GRBs) X-ray afterglows is one 
of the major discovery of the \emph{Swift} satellite. Flares are observed as episodic, large-scale 
amplitude variations in the light-curves with a typical $\Delta t/t \approx 0.1$. Their temporal properties 
(see \cite{Chincarini10}, \cite{Margutti10} and \cite{Bernardini10} for a recent
compilation) make it difficult to interpret the observed emission in the framework of the external shock scenario. 
Instead, the flare emission is likely to be associated to episodes of internal engine activity due to the same source
which powers the gamma-ray prompt pulses. It is therefore of particular interest to investigate the flares-prompt pulses 
connection from the observational point of view: do flares follow the entire set of empirical relations 
found from the analysis of prompt emission pulses? 

\begin{figure}
\label{Fig:060904B}
  \includegraphics[scale=0.7]{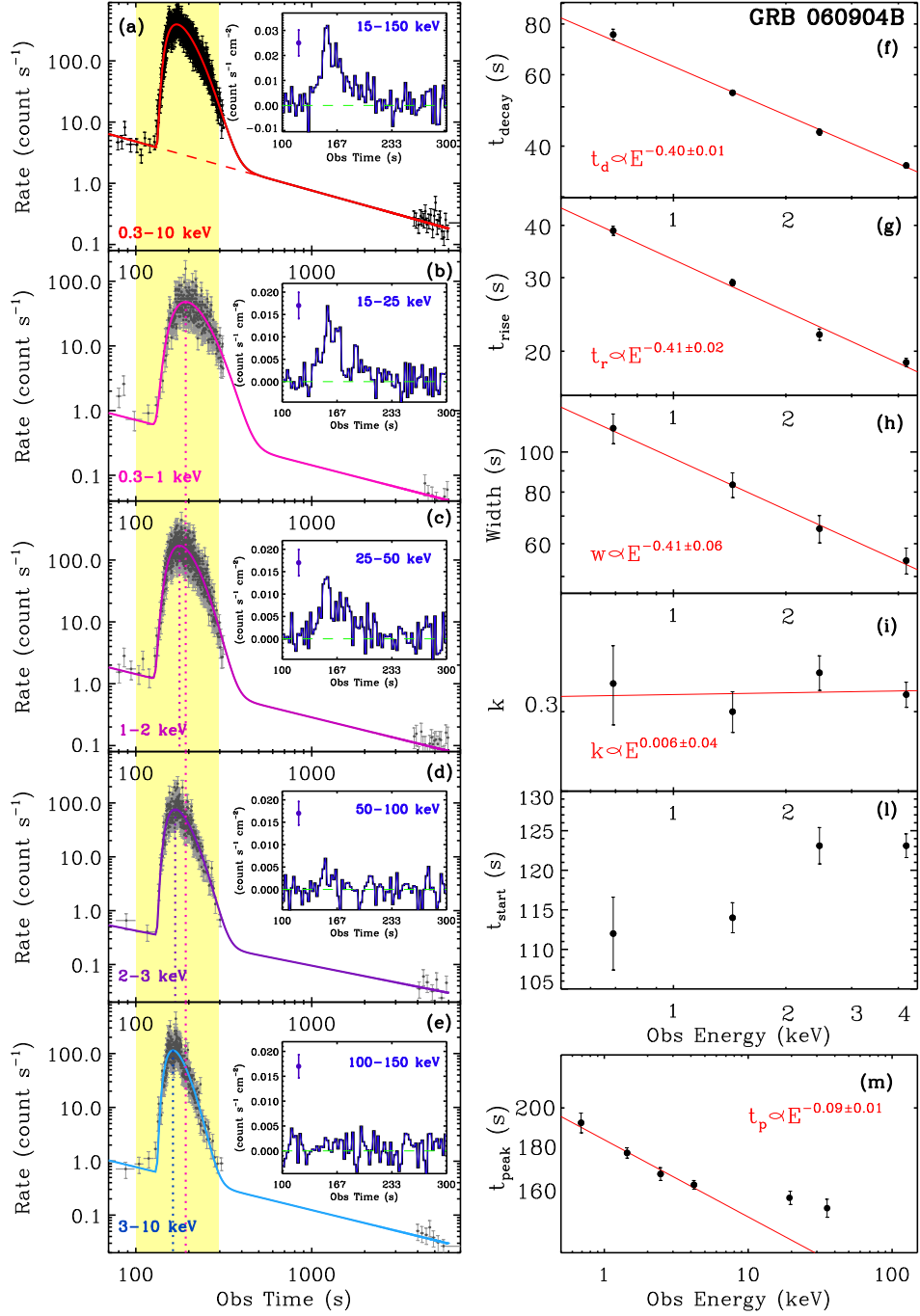}
  \caption{Panels \emph{(a)} through \emph{(e)}: GRB\,060904B flare best fit profile in the total 0.3-10 keV
        XRT energy band and in the 4 channels. \emph{Insets}: BAT signal contemporaneous to the flare emission detected 
        in the 15-150 keV, 15-25 keV,
      25-50 keV, 50-100 keV and 100-150 keV energy bands. The typical uncertainty affecting the BAT data is also
      shown. Panels \emph{(f)} through \emph{(m)}: observed evolution of the flare decay time \emph{(f)}, rise time \emph{(g)},
      width \emph{(h)}, asymmetry \emph{(i)}, start-time \emph{(l)} and peak time \emph{(m)} with observed energy band.
      The best fit power-law relation describing the evolution of each parameter is drawn with a red solid line and explicitly
      written in each panel. The fit reported in panel \emph{(m)} concerns the XRT data only.}
\end{figure}

I address this question analysing a set of 9 bright X-ray flares in 4 different X-ray energy band inside the 
0.3-10 keV of the X-ray telescope (XRT) on board \emph{Swift}. Each flare profile in each energy band is 
modelled using a Norris 2005 profile \cite{Norris05}. Figure \ref{Fig:060904B} shows the result for the flare
detected in GRB\,060904B taken as an example: a general trend can be seen for high energy flare profiles
to rise faster, decay faster and peak before the low energy emission.
The same behaviour is found in the case of prompt pulses (see e.g. \cite{Norris05}). In particular, from Fig.
\ref{Fig:060904B} a tendency is apparent for the peak in the harder channel to lead that in the softer
channel: this directly translates into a measurable, positive \emph{peak lag}.
In the following I concentrate on the analysis of the lag-luminosity relation in long GRB X-ray flares:
I refer the reader to reference \cite{Margutti10} for a detailed and comprehensive analysis of the
temporal and spectral properties of bright X-ray flares.

%------------------------------------------------------------------------------------------
\section{Results and discussion}
\begin{figure}
\label{Fig:laglum}
  \includegraphics[scale=0.65]{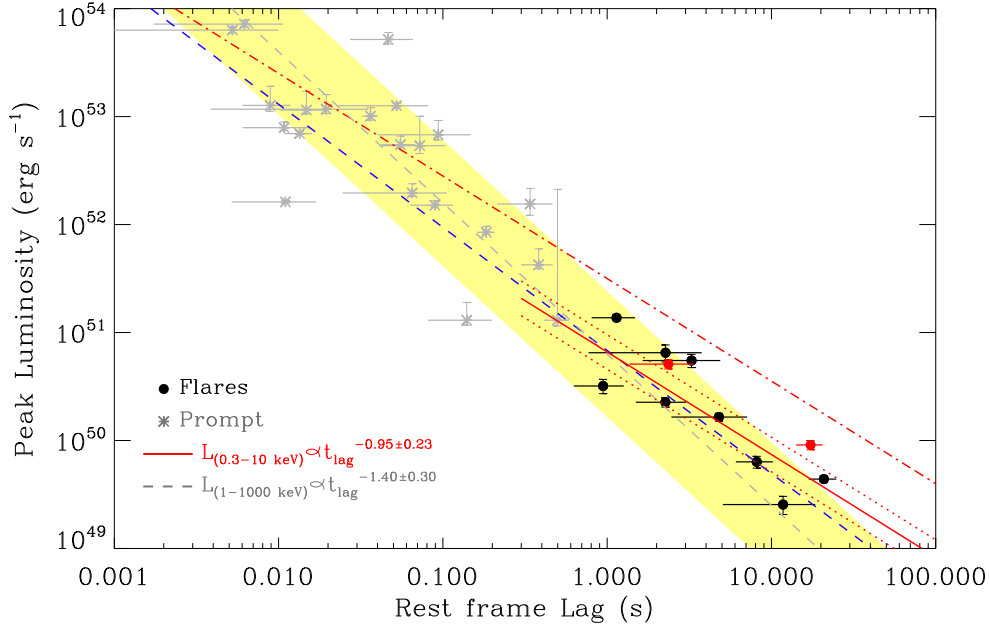}
  \caption{Flare peak-lag peak-luminosity relation. Black filled circles: flares from \cite{Chincarini10}. 
  The isotropic peak luminosity has been computed in the observed 0.3-10 keV energy band. 
  The two flares (in GRB\,060418 and GRB\,060904B) for which it was possible to estimate the 
  Band parameters are marked with red bullets: in these
  cases the 1-10000 keV peak luminosity is plotted. In both cases the flare peak lag is reported.
  Red solid line: flare best fit peak-lag peak-luminosity relation: errors are provided at 90\% c.l. 
  Red dotted lines: best fit sample variance. Grey stars: prompt gamma-ray data corresponding to the
  gold and silver samples of long GRBs from \cite{Ukwatta10}. The isotropic peak luminosity has 
  been calculated in the 1-10000 keV frame. 
  The lag corresponds to the time difference between light-curve structures in the 50-100 keV and 100-200 
  keV channels. Grey dashed line: 
  prompt best fit peak-lag peak-luminosity relation from \cite{Ukwatta10}. The shaded area marks the 
  68\% region around this fit. Red dot-dashed line: best fit relation of the
  gold sample performed accounting for the sample variance %(D'Agostini 2005): $L_{\rm{p,iso}}\propto t_{\rm{lag}}^{-0.95\pm0.30}$.
  Blue dashed line: lag luminosity relation from \cite{Norris2000}.}
\end{figure}

X-ray flares define a rather tight lag-luminosity relation:
\begin{equation}
\frac{L_{\rm{p,iso}}^{\rm{0.3-10\,keV}}}{\rm{(erg\,s^{-1})}}=10^{(50.82\pm0.20)}\times\Big(\frac{t_{\rm{lag,x}}}{\rm{s}}\Big)^{(-0.95\pm0.23)}
\end{equation}
where $t_{\rm{lag,x}}$ is the rest frame peak lag calculated as the difference in time between the flare peak times in the 
0.3-1 keV and the 3-10 keV energy bands. The subscript $x$ reminds that the time lag is calculated in the X-ray regime.
We properly account for the sample variance following the method outlined by \cite{Dagostini05}.
Errors are provided at 90\% c.l.
%It is not possible to constrain the spectral peak energy and the spectral slopes of the majority of the flares of C10: for this reason 
%we fitted the spectrum extracted around the peak time with an absorbed SPL within \textsc{XSPEC} and conservatively quote 
%the isotropic peak luminosity
%as obtained in the observed 0.3-10 keV bandpass (black bullets in Fig. \ref{Fig:laglum}). In two cases, for GRB\,060418 and 
%GRB\,060904B the Band parameters were determined thanks to the BAT+XRT coverage (Table \ref{Tab:spectot}): this allows us to 
%plot the 1-10000 keV (rest frame) isotropic peak luminosity. However, the best fit relation was derived using their 0.3-10 keV
%values for homogeneity (the two red bullets in Fig. \ref{Fig:laglum} consistently lie above the expectation). In the cases
%of the GRB\,060418 and GRB\,090904B flares, the isotropic peak luminosity in the 0.3-10 keV observed band underestimates the
%1-10000 keV rest frame value of a factor 2-3.

The lag-luminosity is one of the key relations which connects the GRB \emph{prompt} temporal and spectral properties: discovered by 
\cite{Norris2000} as a time integrated property of each particular burst, the relation was conclusively demonstrated  
to reflect \emph{pulses} rather than \emph{bursts} properties by \cite{Hakkila08}. Figure \ref{Fig:laglum} shows a direct comparison
between the flare and the prompt properties in the lag-luminosity diagram: this is of particular interest since long and short bursts
are known to occupy different regions of the plane (see e.g. \cite{Gehrels06}). Flares in long GRB are consistent with the long GRBs
lag-luminosity relation. However, we should consider that: first, in Fig. \ref{Fig:laglum} the lag of the prompt data is calculated 
using the Cross Correlation Function (CCF) to the entire BAT light-curve and consequently reflects a time integrated property 
\cite{Ukwatta10}. Second: for the prompt data the lag is defined as the time difference between light-curve structures 
in the 50-100 keV and 100-200 keV channels. For the flares we calculated the \emph{peak} lag between X-ray energy bands. 
In both cases the lag has been computed between band-passes around the event spectral peak energy. Third: the prompt peak luminosity is
calculated in the 1-10000 keV rest frame energy band, while for the flares the peak luminosity is calculated from the 0.3-10 keV
(observed) bandpass (which is expected to be a factor of 2-3 below the 1-10000 keV value).

With these caveats in mind, it is remarkable that the best fit slope of the flare lag-luminosity is consistent with the
results based on BAT data and reported in \cite{Ukwatta10}: %accounting for the sample  variance following \cite{Dagostini05}, the best fit relation derived
%from the gold sample reads: 
$L_{\rm{p,iso}}\propto t_{\rm{lag},\gamma}^{(-0.95\pm0.30)}$ (red dot dashed line in Fig. \ref{Fig:laglum}).
It is not surprising that this relation overestimates the flare luminosity which is computed in a narrower energy window by a factor $\sim 5$.
Only a marginal consistency can be quoted with \cite{Hakkila08} who reported an index $\sim0.6$.
Our findings are instead fully consistent with \cite{Norris2000}  (both in normalization and index), with a power-law 
index 1.14 (blue dashed line in Fig. \ref{Fig:laglum}): these authors reported lags between BATSE energy bands 100-300 kev 
and 25-50 keV. The same is true if we consider the lag-luminosity power-law index by \cite{Schaefer07} who reported a value of 1.01. 
%The prompt lag-luminosity relations are summarized in Table \ref{Tab:laglum}.

This result strongly suggests that whatever is the mechanism at work in the GRB prompt emission, the same is also the source
of the X-ray flare emission hundreds of seconds later. The lag-luminosity relation has been proven to be a fundamental law extending $\sim5$ 
decades in energy and $\sim5$ decades in time.

%%%%%%%%%%%%%%%%%%%%%%%%%%%%%%%%%%%%%%%%%%%%%%%%
%% BACKMATTER
%%%%%%%%%%%%%%%%%%%%%%%%%%%%%%%%%%%%%%%%%%%%%%%%

%\begin{theacknowledgments}
  
%\end{theacknowledgments}

\bibliographystyle{aipproc}   % if natbib is available

\bibliography{Margutti}

\end{document}